\documentclass[emulateapj]{emulateapj}
\usepackage{amsmath}
\usepackage{lineno}

\newcommand{\teff}{\mbox{$T_{\rm eff}$}}
\newcommand{\logg}{\mbox{$\log g$}}
\newcommand{\feh}{\mbox{$\rm{[Fe/H]}$}}
\newcommand{\fehval}{\mbox{$+0.42$}}

\newcommand{\numax}{\mbox{$\nu_{\rm max}$}}
\newcommand{\dnu}{\mbox{$\Delta \nu$}}
\newcommand{\muHz}{\mbox{$\mu$Hz}}
\newcommand{\thestar}{\mbox{K2-97}}
\newcommand{\planrad}{\mbox{1.31 $\pm$ 0.11 R$_{\mathrm{J}}$}}
\newcommand{\planmass}{\mbox{1.10 $\pm$ 0.11 M$_{\mathrm{J}}$}}
\newcommand{\starrad}{\mbox{4.20 $\pm$ 0.14 R$_{\odot}$}}
\newcommand{\starmass}{\mbox{1.16 $\pm$ 0.12 M$_{\odot}$}}

\begin{document}



\bibliographystyle{apj} 


\title{K2-97\lowercase{b}: A (Re-?)Inflated Planet Orbiting a Red Giant Star}
\author{Samuel K.\ Grunblatt\altaffilmark{1}}

\email{Email: skg@ifa.hawaii.edu}
\shorttitle{The First Re-Inflated Planet}
\shortauthors{Grunblatt et al.}

\author{Daniel Huber\altaffilmark{2,3,4}}
\author{Eric J.\ Gaidos\altaffilmark{5,6}}
\author{Eric D.\ Lopez\altaffilmark{7}}
\author{Benjamin J.\ Fulton\altaffilmark{1,14}}
\author{Andrew Vanderburg\altaffilmark{8,14}}
\author{Thomas Barclay\altaffilmark{9}}
\author{Jonathan J.\ Fortney\altaffilmark{10}}
\author{Andrew W.\ Howard\altaffilmark{1,11}}
\author{Howard T. Isaacson\altaffilmark{12}}
\author{Andrew W.\ Mann\altaffilmark{13,15}}
\author{Erik Petigura\altaffilmark{11,15}}
\author{Victor Silva Aguirre\altaffilmark{4}}
\author{Evan J.\ Sinukoff\altaffilmark{1,16}}

\altaffiltext{1}{Institute for Astronomy, University of Hawaii,
2680 Woodlawn Drive, Honolulu, HI 96822, USA}
\altaffiltext{2}{Sydney Institute for Astronomy (SIfA), School of Physics, University of 
Sydney, NSW 2006, Australia}
\altaffiltext{3}{SETI Institute, 189 Bernardo Avenue, Mountain View, CA 94043, USA}
\altaffiltext{4}{Stellar Astrophysics Centre, Department of Physics and Astronomy, 
Aarhus University, Ny Munkegade 120, DK-8000 Aarhus C, Denmark}
\altaffiltext{5}{Department of Geology $\&$ Geophysics, University of
Hawaii at Manoa, Honolulu, Hawaii 96822, USA} 
\altaffiltext{6}{Center for Space and Habitability, University of Bern, Bern, Switzerland CH-3012}
\altaffiltext{7}{Institute for Astronomy, Royal Observatory Edinburgh, University of Edinburgh, Blackford Hill, Edinburgh, UK}
\altaffiltext{8}{HarvardÐSmithsonian Center for Astrophysics, 60 Garden St., Cambridge, MA 02138, USA}
\altaffiltext{9}{NASA Ames Research Center, Moffett Field, CA 94035, USA}
\altaffiltext{10}{Department of Astronomy and Astrophysics, University of California, Santa Cruz, CA 95064, USA}
\altaffiltext{11}{California Institute of Technology, Pasadena, CA 91125, USA}
\altaffiltext{12}{Department of Astronomy, UC Berkeley, Berkeley, CA 94720, USA}
\altaffiltext{13}{Department of Astronomy, The University of Texas at Austin, Austin, TX 78712, USA}
\altaffiltext{14}{National Science Foundation Graduate Research Fellow}
\altaffiltext{15}{Hubble Fellow}
\altaffiltext{16}{Natural Sciences and Engineering Research Council of Canada Graduate Student Fellow}

\begin{abstract}

Strongly irradiated giant planets are observed to have radii larger than thermal evolution models predict. Although these inflated planets have been known for over fifteen years, it is unclear whether their inflation is caused by deposition of energy from the host star, or inhibited cooling of the planet. These processes can be distinguished if the planet becomes highly irradiated only when the host star evolves onto the red giant branch. We report the discovery of K2-97b, a \planrad, \planmass{} planet orbiting a \starrad, \starmass{} red giant star with an orbital period of 8.4 days. We precisely constrained stellar and planetary parameters by combining asteroseismology, spectroscopy, and granulation noise modeling along with transit and radial velocity measurements. The uncertainty in planet radius is dominated by systematic differences in transit depth, which we measure to be up to 30\% between different lightcurve reduction methods. Our calculations indicate the incident flux on this planet was 170$^{+140}_{-60}$ times the incident flux on Earth while the star was on the main sequence. Previous studies suggest that this incident flux is insufficient to delay planetary cooling enough to explain the present planet radius. This system thus provides the first evidence that planets may be inflated directly by incident stellar radiation rather than by delayed loss of heat from formation. Further studies of planets around red giant branch stars will confirm or contradict this hypothesis, and may reveal a new class of re-inflated planets.





\end{abstract}

\section{Introduction}

The first measurements of the radius of a planet outside our solar system were reported by \citet{charbonneau2000} and \citet{henry2000}. These groundbreaking measurements also revealed a mystery in exoplanet science: the planet radius was considerably larger than expected from planet models \citep{burrows1997, bodenheimer2001, guillot2002}. Further transit studies of giant planets in short period orbits revealed similarly enlarged planets \citep{cameron1999, hebb2009}. Although very young ($<$ 10 Myr) planets are expected to have large radii ($>$1.2 R$_{\mathrm{J}}$) due to heat from formation, this cannot explain the dozens of known planets with radii $>$1.2 R$_{\mathrm{J}}$ orbiting several billion year old stars \citep{guillot2014}. Moreover, a correlation has been observed between incident stellar radiation and planetary radius inflation \citep{burrows2000, laughlin2011, lopez2016}. 


Several potential mechanisms for planet inflation have been suggested \citep{baraffe2014}, but these mechanisms can generally be placed into two broad classes. In the first class, $\lesssim$1\% of the stellar irradiance is deposited into the planet's interior, causing the planet to heat and expand \citep{batygin2010}. In the second class, the planet retains its initial heat from formation and remains inflated due to stalled contraction \citep{chabrier2007, wu2013}. A planet with an orbital period of $\sim$10-30 days would be too cool to be inflated around a solar-type main sequence star, but would experience irradiation $>$500 times the flux on Earth for more than 100 Myr while its host star evolves onto the red giant branch. Thus, the discovery of an inflated planet in this period range around an evolved star would indicate that inflation is a response to high stellar irradiation, whereas a population of exclusively non-inflated gas giant planets would suggest that inflation is governed more strongly by delayed cooling \citep{lopez2016}.


Searches for planets around evolved stars may also provide clues to understanding the occurrence of planets around stars more massive than the Sun. Massive stars have been observed to produce more giant planets than small stars \citep{johnson2009, gaidos2013}, suggesting that these stars have more planet-forming material than small stars \citep{andrews2013}. However, the larger radii of these stars make planet transit signals smaller. More importantly, the fast rotation and relatively few absorption lines of main sequence, intermediate-mass ($\geq$1.5 M$_\odot$) stars made planet detection using radial velocities difficult before the \emph{Kepler} era. However, these F- and A-type stars evolve into G- and K-type giants with deeper absorption lines and slower rotation rates, allowing precise radial velocity measurement. Early radial velocity surveys to investigate planet occurrence as a function of stellar mass included evolved stars \citep{johnson2007b}, and indicated a strong correlation between planet occurrence and stellar mass. However, this correlation is heavily debated, as the short lives and intrinsic rarity of these stars result in systematic uncertainties on host star masses derived from stellar models \citep{lloyd2011, schlaufman2013, lloyd2013, johnson2013, johnson2014, ghezzi2015}.

To answer the questions of giant planet occurrence and inflation, we have begun a search for transiting planets orbiting giant stars with the NASA K2 Mission \citep{howell2014, huber2015}. By targeting low-luminosity red-giant branch (RGB) stars which oscillate with frequencies detectable with K2's long-cadence data, stellar radius and mass can be precisely determined using asteroseismology for stars around which giant planet transits are detectable. This precision is crucial to investigate the mechanisms for planet inflation and the dependence of planet occurrence on stellar mass. Here, we present the discovery and characterization of the first planet from our survey.

\section{Observations}

\subsection{K2 Photometry}
 
 In the K2 extension to the NASA \emph{Kepler} mission, multiple fields along the ecliptic are observed almost continuously for approximately 80 days \citep{howell2014}. EPIC 211351816 (now known as K2-97) was selected for observation as a part of K2 Guest Observer Proposal GO5089 (PI: Huber) and observed in Campaign 5 of K2 during the first half of 2015.  As the \emph{Kepler} telescope now has unstable pointing due to the failure of two of its reaction wheels, it is necessary to correct for the pointing-dependent error in the flux received per pixel. We produced a lightcurve by simultaneously fitting thruster systematics, low frequency variability, and planet transits with a Levenberg-Marquardt minimization algorithm, using a modified version of the pipeline from \citet{vanderburg2016}.
  
\begin{figure}[ht!]
\epsscale{1.15}
\plotone{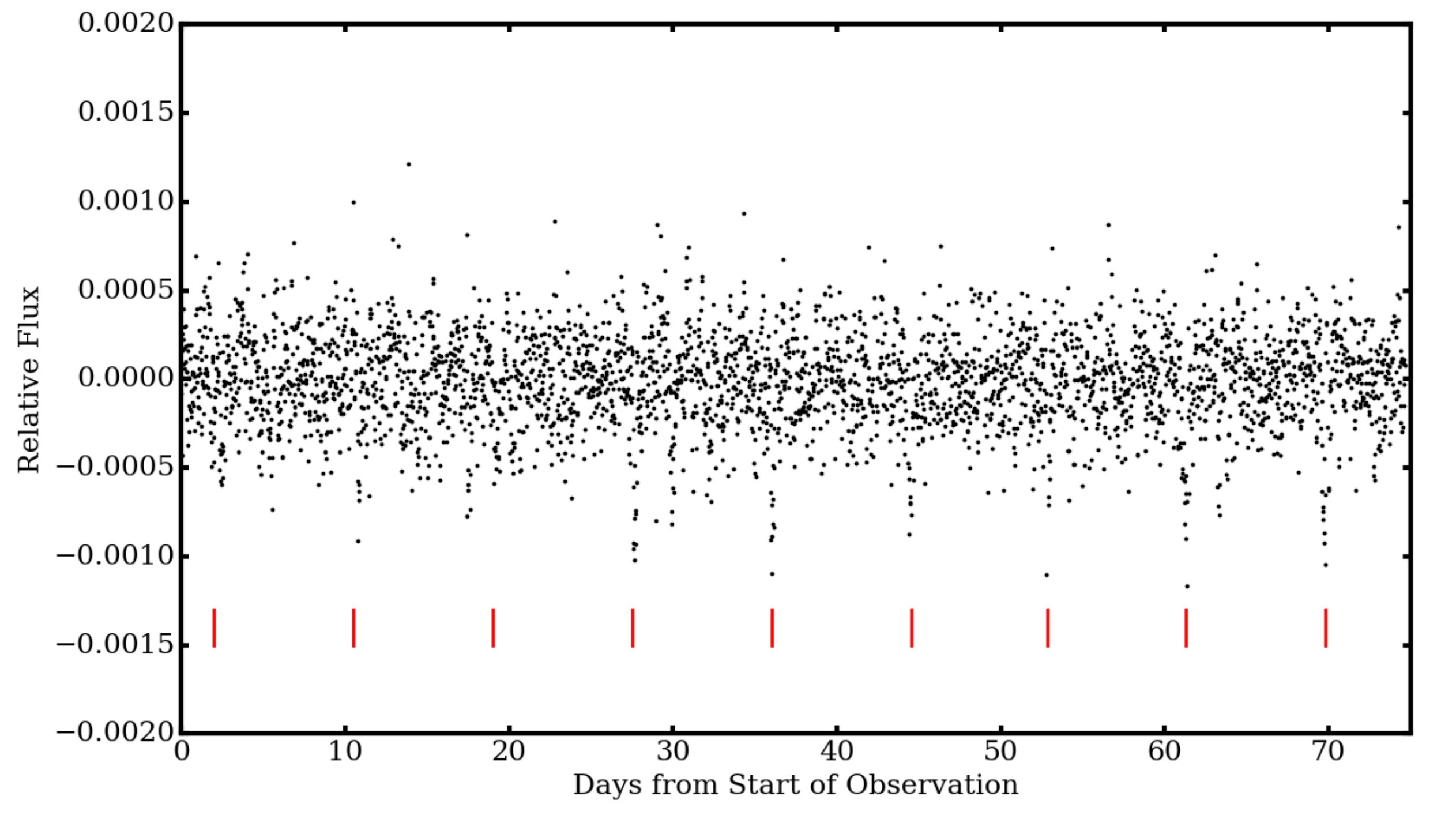}
\caption{Detrended K2 lightcurve of K2-97. This lightcurve was produced using a modified method of the pipeline presented in \citet{vanderburg2016}, where both instrument systematics and the planet transit were modeled simultaneously to prevent transit dilution. The lightcurve has been normalized as well as unity subtracted. Individual transits are visible by eye, and are denoted by red fiducial marks. \label{lightcurve}}
\end{figure}

We also analyzed the PDC-MAP light curve provided by the K2 Science Office \citep{stumpe2012,smith2012} as well as the detrended lightcurves created with the methods of \citet{vanderburg2016}, \citet{petigura2015}, and \citet{aigrain2016}. The use of different lightcurves resulted in statistically significant differences in the transit depth, illustrating the additional systematic uncertainties introduced by lightcurve reductions (see $\S$ 5.1 for more details). However, the results from all lightcurves analyzed were broadly consistent with the modified \citet{vanderburg2016} results (see Discussion). Figure \ref{lightcurve} shows our adopted lightcurve for K2-97.



\subsection{Imaging with Keck/NIRC2 AO}

Natural guide-star adaptive optics (AO) images of \thestar{} were obtained through the broad K$'$ filter ($\lambda_{\mathrm{center}}$ = 2.124 $\mu$m) with the Near-Infrared Camera (NIRC2) at the Keck-2 telescope on Mauna Kea during the nights of UT 19 March and 12 May 2016. The narrow camera (pixel scale 0.01") was used for both sets of observations.  No additional sources were detected within $\sim$3"  of the star.  The contrast ratio of the detection limit is more than 7 magnitudes at 0.5"; brighter objects could be detected to within 0.15" of the star.

\subsection{Spectroscopy with UH88/SNIFS, IRTF/SpeX, and Keck/HIRES}

We obtained a high resolution, high signal-to-noise spectrum of \thestar{} using the High Resolution Echelle Spectrometer (HIRES) on the 10 meter Keck-I telescope at Mauna Kea Observatory on the Big Island of Hawaii. HIRES provides spectral resolution of roughly 100,000 in a wavelength range of 0.3 to 1.0 microns \citep{vogt1994}. We also obtained medium-resolution optical and infrared spectra using the Supernova Integrated Field Spectrograph (SNIFS) on the 2.2 meter University of Hawaii telescope and SpeX on the 3 meter Infrared Telescope Facility (IRTF), providing spectral resolution of 1000--2000 over a wavelength range from 0.3 to 5.5 microns \citep{lantz2004, rayner2003}.

\begin{figure}[ht!]
\epsscale{1.0}
\plotone{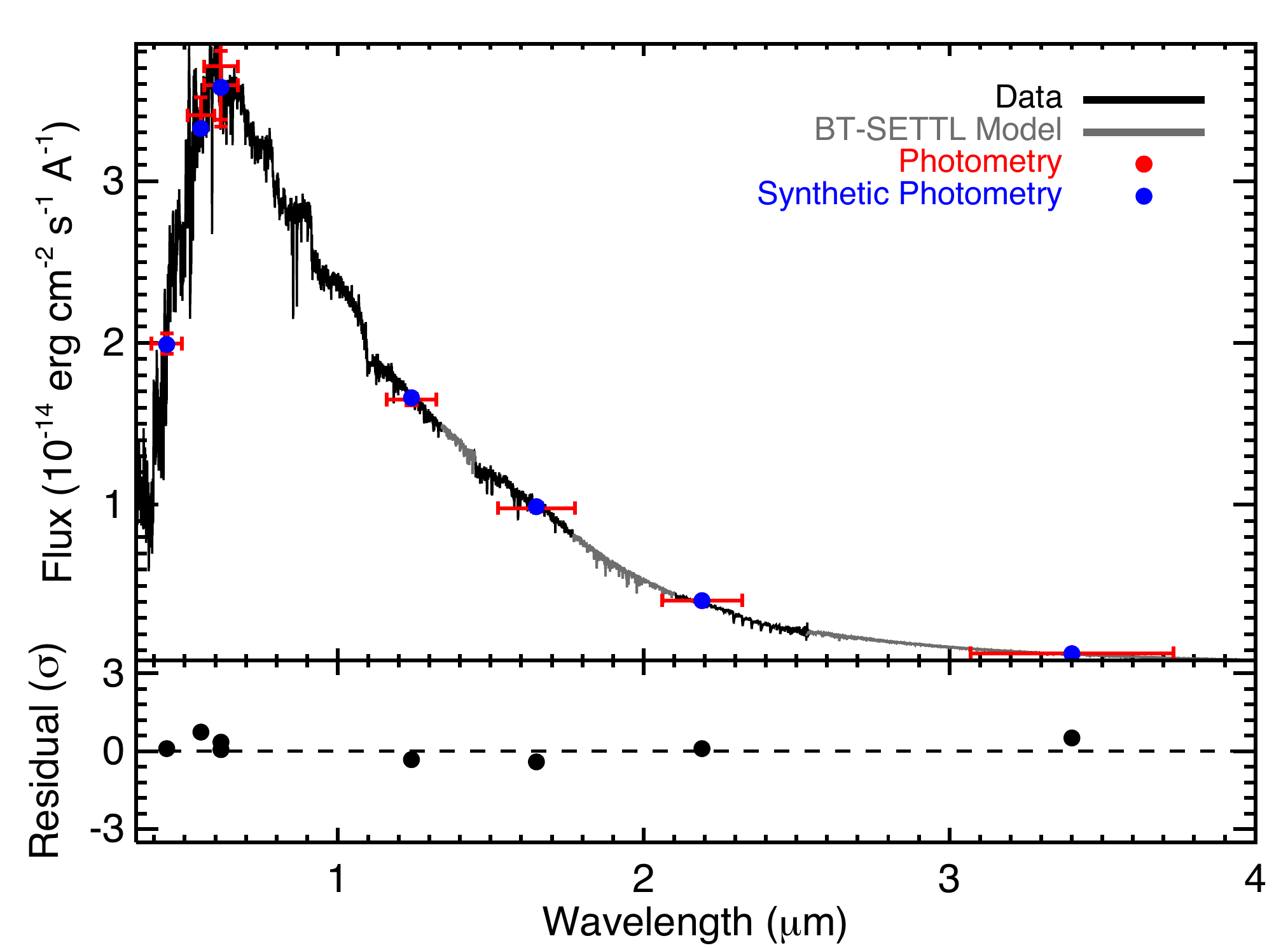}
\caption{Flux calibrated optical and NIR spectrum of EPIC 211351816. Photometry is shown in red, with the horizontal error bars representing the effective width of the filter. Synthetic photometry derived from the spectrum is shown in blue. We replaced regions of high telluric absorption and those outside the range of our empirical spectra with an atmospheric model, which we show in grey. The spectrum and photometry shown here have not been corrected for reddening. The bottom panel shows the residual (photometry-synthetic) in units of standard deviations.
\label{spectrum}}
\end{figure}

 
 We joined and flux calibrated the SNIFS and SpeX spectra following the method outlined in \citet{Mann2015b}. We first downloaded photometry from the Two-Micron All-Sky Survey \citep[2MASS,][]{Skrutskie2006}, AAVSO All-Sky Photometric Survey \citep[APASS,][]{Henden2012}, and The Wide-field Infrared Survey Explorer \citep[WISE,][]{Wright2010}. The spectrum and all photometry were converted to physical fluxes using the appropriate zero-points and filter profiles \citep{cohen2003,Jarrett2011,Mann2015a}. We scaled the optical and NIR spectra to match the photometry and each other in overlapping regions (0.8-0.95 $\mu$m), accounting for correlated errors in the flux calibration. Regions of high telluric contamination or missing from our spectrum (e.g., beyond 2.4 $\mu$m) were replaced with a best-fit atmospheric model from the BT-SETTL grid \citep{Allard2011, Allard2013}. The final calibrated and combined spectrum is shown in Figure~\ref{spectrum}.

\subsection{Radial Velocity Measurements}

Radial velocity measurements were obtained between January 27 and May 16, 2016 using the High Resolution Echelle Spectrometer (HIRES) on the Keck-I Telescope at the Mauna Kea Observatory in Hawaii and the Levy spectrometer on the Automated Planet Finder (APF) telescope at Lick Observatory in California. The specific measurements are listed in Table \ref{tbl-rv}. The nine spectra observed were obtained using an iodine cell. Measurements with the Keck telescope achieved a precision of greater than 1 m s$^{-1}$, whereas the APF measurements have measurement uncertainties of $\sim$30 m s$^{-1}$. We collected three measurements with Keck/HIRES and six with APF.

The Levy Spectrograph is a high-resolution slit-fed optical echelle spectrograph mounted at one of the two Nasmyth foci of the APF designed specifically for the detection and characterization of exoplanets \citep{burt2014, fulton2015}. Each spectrum covers a continuous wavelength range from 3740 to 9700 \AA. We observed EPIC 211351816 using a 1.0" wide decker for an approximate spectral resolution of R = 100,000. Starlight passed through a cell of gaseous iodine which serves as a simultaneous calibration source for the instrumental PSF and wavelength reference. We measured relative RVs using a Doppler pipeline descended from the iodine technique in \citet{butler1996}. We forward-modeled 848 segments of each spectrum between 5000 and 6200 \AA. The model consists of a stellar template spectrum, an ultra high-resolution Fourier transform spectrum of the iodine absorption of the Levy cell, a spatially variable PSF, a wavelength solution, and RV. Traditionally, a high signal-to-noise iodine-free observation of the same star is deconvolved with the instrumental PSF and used as the stellar template in the forward modeling process. However, in this case the star is too faint to collect the signal-to-noise needed for reliable deconvolution in a reasonable amount of time on the APF. Instead, we simulated this observation by using the SpecMatch software \citep{petigura2015} to construct a synthetic template from the \citet{coelho2014} models and best-fit stellar parameters.

\begin{deluxetable}{cccrrrr}
\tabletypesize{\scriptsize}
\tablecaption{Radial Velocities\label{tbl-rv}}
\tablewidth{0pt}
\tablehead{
\colhead{BJD-2440000} & \colhead{RV (m s$^{-1}$)} & \colhead{Prec. (m s$^{-1}$)} & \colhead{Tel./inst. used}
}
\startdata
17414.927751 & 14.84 & 0.68 & Keck/HIRES\\
17422.855362 & -17.18 & 0.72 & Keck/HIRES\\
17439.964043 & 1.92 & 0.82 & Keck/HIRES\\
17495.743272 & -2 & 24 & APF/Levy\\
17498.729824 & -30 & 27 & APF/Levy\\
17505.670536 & -84 & 39 & APF/Levy\\
17507.723056 & 27 & 30 & APF/Levy\\
17524.687701 & 0 & 32 & APF/Levy\\
17525.686520 & 67 & 30 & APF/Levy\\

\enddata 
\tablecomments{The precisions listed here are instrumental only, and do not take into account the uncertainty introduced by stellar jitter. For evolved stars, radial velocity jitter on relevant timescales is typically $\sim$5 m s$^{-1}$ (see text).}

\end{deluxetable}

\section{Host Star Characteristics}

\subsection{Spectroscopic Analysis}



In order to obtain precise values for the stellar parameters, we collected a moderate signal-to-noise iodine-free observation using the HIRES spectrograph on the Keck I telescope \citep{vogt1994}. We measured the effective temperature (\teff), surface gravity (\logg), iron abundance (\feh), and rotational velocity of the star using the tools available in the SpecMatch software package \citep{petigura2015}. We first corrected the observed wavelengths to be in the observerÕs rest frame by cross-correlating a solar model with the observed spectrum. Then we fit for \teff, \logg, \feh{}, $v$sin$i$, and the instrumental PSF using the underlying Bayesian differential-evolution Markov Chain Monte Carlo machinery of ExoPy \citep{fulton2013}. At each step in the MCMC chains, a synthetic spectrum is created by interpolating the \citet{coelho2014} grid of stellar models for a set of \teff, \logg, and \feh{} values and solar alpha abundance. We convolved this synthetic spectrum with a rotational plus macroturbulence broadening kernel using the prescriptions of \citet{valenti2005} and \citet{hirano2011}. Finally, we performed another convolution with a Gaussian kernel to account for the instrumental PSF, and compared the synthetic spectrum with the observed spectrum to assess the goodness of fit. The priors are uniform in \teff, \logg, and \feh{} but we assign a Gaussian prior to the instrumental PSF that encompasses the typical variability in the PSF width caused by seeing changes and guiding errors. Five echelle orders of the spectrum were fit separately and the resulting posterior distributions were combined before taking the median values for each parameter. Parameter uncertainties were estimated as the scatter in spectroscopic parameters given by SpecMatch relative to the values for 352 stars in the in \citet{valenti2005} sample and 76 stars in the \citet{huber2013} asteroseismic sample. Systematic trends in SpecMatch values as a function of \teff, \logg, and \feh{} relative to these benchmark samples were fit for and removed in the final quoted parameter values. Initial fits to the stellar spectrum for  \teff, \logg, \feh{}, and $v$sin$i$ were made without asteroseismic constraints, and were found to be in good agreement with the asteroseismic quantities. A prior was applied to the value for \logg{} based on the asteroseismic estimate of 3.26 $\pm$ 0.015 (see Section 3.2), which resulted in convergence to the values listed in Table \ref{tbl-star}.

\subsection{Asteroseismology}

\begin{figure}
\begin{center}
\resizebox{\hsize}{!}{\includegraphics{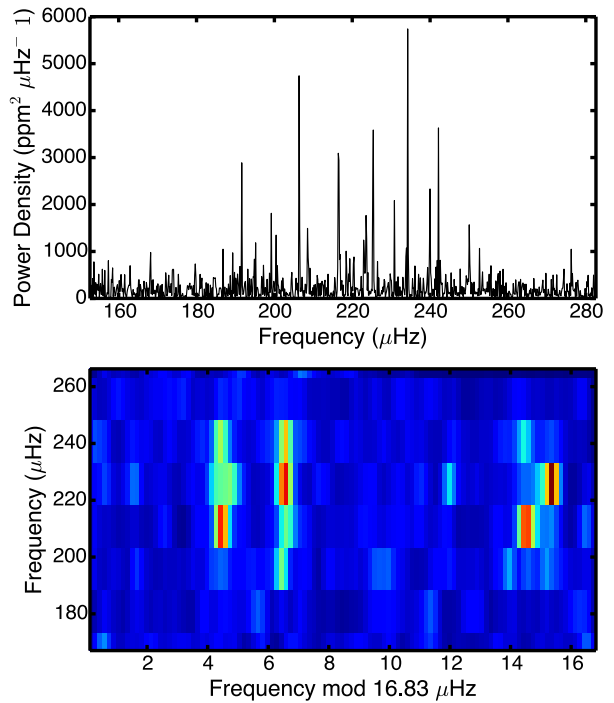}}
\caption{Top panel: Power spectrum of the K2 time series centered on the frequency region with detected oscillations. Bottom panel: Echelle diagram of the granulation background-corrected power spectrum using $\Delta \nu =16.83 \mu$\,Hz. Oscillation modes with $l=0,2$ (left) and $l=1$ (right) are visible. Note that dipole mode series is more complex due to the presence of mixed modes.}
\label{seismo}
\end{center}
\end{figure}

 Stellar oscillations are a powerful tool to determine precise fundamental properties of exoplanet host stars \citep[e.g][]{cd10,gilliland11,huber2013}. The top panel of Figure \ref{seismo} shows the power spectrum calculated from the K2 data after removing the transits from the light curve. We detect a strong power excess with regularly spaced peaks near $\sim$\,220\,\muHz ~(75 minutes), typical for an oscillating low-luminosity red giant star. 

The power excess can be characterized by the frequency of maximum power (\numax) and the average separation of modes with the same spherical degree and consecutive radial order (\dnu). To measure \numax\ and \dnu\ we analyzed the K2SC lightcurve of this system \citep{aigrain2016} using the method of \citet{huber09}, which corrects the background granulation noise by fitting a 2-component Harvey model \citep{harvey85} in the frequency domain. The frequency of maximum power was then measured from the peak of the heavily smoothed, background-corrected power spectrum, and \dnu\ was measured using an autocorrelation of the power spectrum. We calculated uncertainties using 1000 Monte Carlo simulations as described in \citet{huber11}, yielding $\numax = 223.7 \pm 5.4 ~\muHz$ and $\dnu = 16.83 \pm 0.17 ~\muHz$.

The bottom panel of Figure \ref{seismo} shows an \'{e}chelle diagram, which stacks radial orders on top of each other, showing the asymptotic spacing of oscillation modes with the same spherical degree $l$. The \'{e}chelle diagram of \thestar{} shows the characteristic signature of nearly vertically aligned quadrupole ($l=2$) and radial ($l=0$) modes, while the dipole modes ($l=1$) show a more complex distribution due to the coupling of pressure modes with gravity mode in the core \citep[known as mixed modes, e.g.][]{dziembowski01,montalban10,bedding10b}. The position of the $l=0$ ridge agrees with the expected value for a low-luminosity RGB star \citep{huber2010,corsaro12}.

To estimate stellar properties from \numax\ and \dnu, we use the scaling relations of \citet{brown91,kb95}:

\begin{equation}
\frac{\Delta \nu}{\Delta \nu_{\odot}} \approx f_{\Delta \nu} \left(\frac{\rho}{\rho_{\odot}}\right)^{0.5} \: ,
\end{equation}

\begin{equation}
\frac{\nu_{\rm max}}{\nu_{\rm max, \odot}} \approx \frac{g}{{\rm g}_{\odot}} \left(\frac{T_{\rm eff}}{T_{\rm eff, \odot}}\right)^{-0.5} \: .
\end{equation}

\noindent
Equations (1) and (2) can be rearranged to solve for mass and radius:

\begin{equation}
\frac{M}{\rm {\rm M}_\odot}   \approx   \left(\frac{\nu_{\rm
max}}{\nu_{\rm max,
\odot}}\right)^{3}\left(\frac{\Delta \nu}{f_{\Delta \nu}
\Delta \nu_{\odot}}\right)^{-4}\left(\frac{T_{\rm eff}}{T_{\rm
eff, \odot}}\right)^{1.5}   
\end{equation}

\begin{equation}
\frac{R}{\rm R_\odot}   \approx  \left(\frac{\nu_{\rm
max}}{\nu_{\rm max, \odot}}\right)\left(\frac{\Delta
\nu}{f_{\Delta \nu} \Delta \nu_{\odot}}\right)^{-2}\left(\frac{T_{\rm
eff}}{T_{\rm eff, \odot}}\right)^{0.5}.
\end{equation}

Our adopted solar reference values are $\nu_{\rm max, \odot}=3090\,\muHz$ and $\Delta \nu_{\odot}=135.1\,\muHz$ \citep{huber11}, as well as $T_{\rm eff, \odot}=5777$\,K.

\begin{deluxetable*}{ccccccrrrrrrrrcrl}
\tabletypesize{\scriptsize}
\tablecaption{Stellar and Planetary Properties\label{tbl-star}}
\tablewidth{0pt}
\tablehead{
\colhead{Property} & \colhead{Value} & \colhead{Source}
}
\startdata
ID & \thestar{}, EPIC 211351816, 2MASS 08310308+1050513 & \citet{huber2016} \\
\emph{Kepler} Magnitude & 12.409 & \citet{huber2016} \\
T$_{\mathrm{eff}}$ & 4790 $\pm$ 90 K & spectroscopy\\
 V$sin(i)$ &  2.8 $\pm$ 1.6 km s$^{-1}$ & spectroscopy\\
 ~[Fe/H] & \fehval{} $\pm$ 0.08 & spectroscopy\\
Stellar Mass, $M_{\mathrm{star}}$ & \starmass & asteroseismology \\
Stellar Radius, $R_{\mathrm{star}}$ & \starrad & asteroseismology \\
Density, $\rho_{*}$ & 0.0222 $\pm$ 0.0004 g cm$^{-3}$ & asteroseismology\\
log $g$ & 3.26 $\pm$ 0.01 & asteroseismology \\
Age & 7.8 $\pm$ 2 Gyr & isochrones\\

\hline
\hline

Planet Radius, R$_{\mathrm{p}}$ & \planrad & asteroseismology, GP+transit model \\
Orbital Period $P_{\mathrm{orb}}$ & 8.4061 $\pm$ 0.0015 days & GP+transit model\\ 
Planet Mass, M$_{\mathrm{p}}$ &  \planmass & asteroseismology, RV model \\
\enddata 
\end{deluxetable*}


\begin{figure*}[ht!]
\epsscale{1.0}
\plottwo{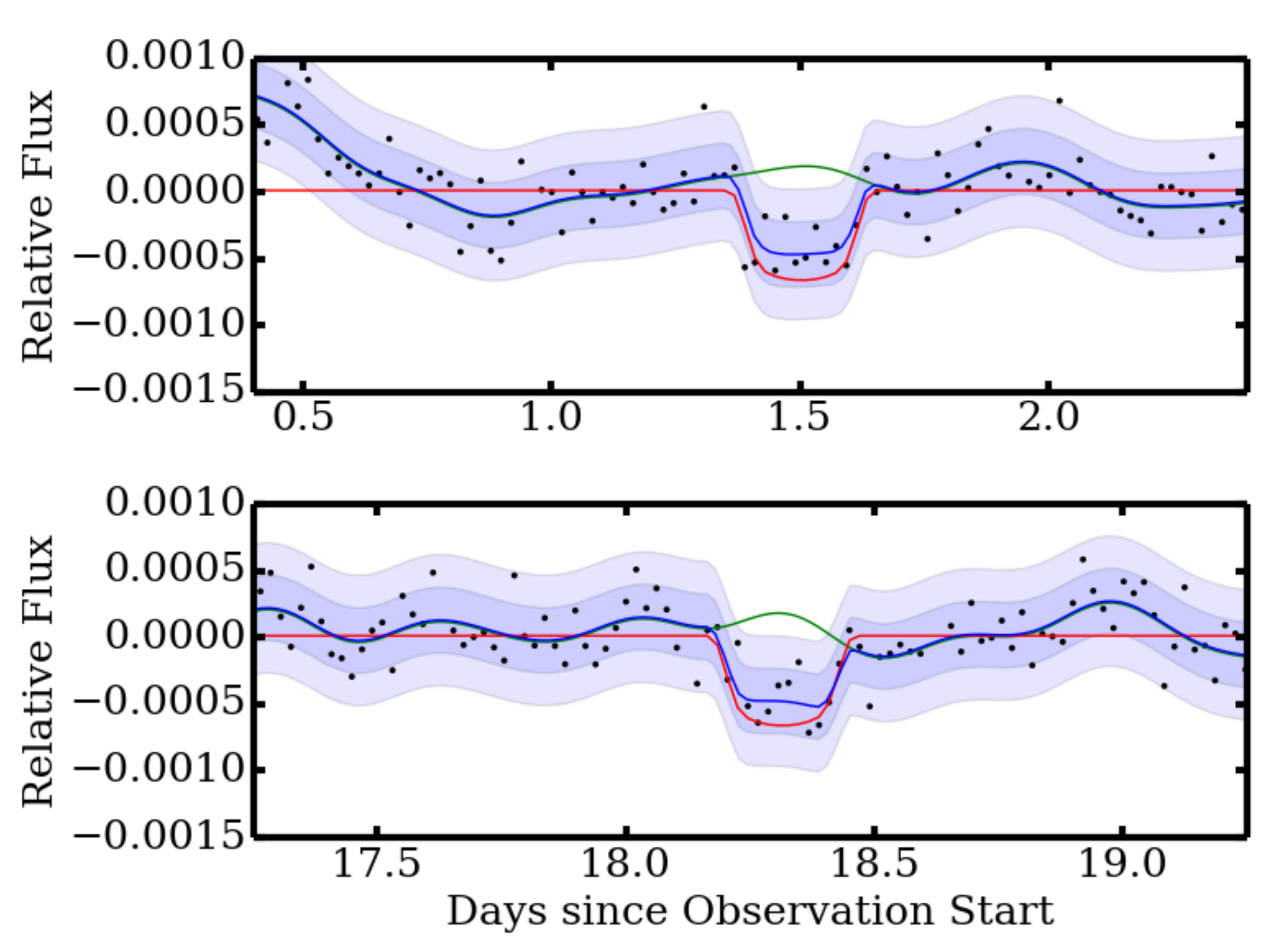}{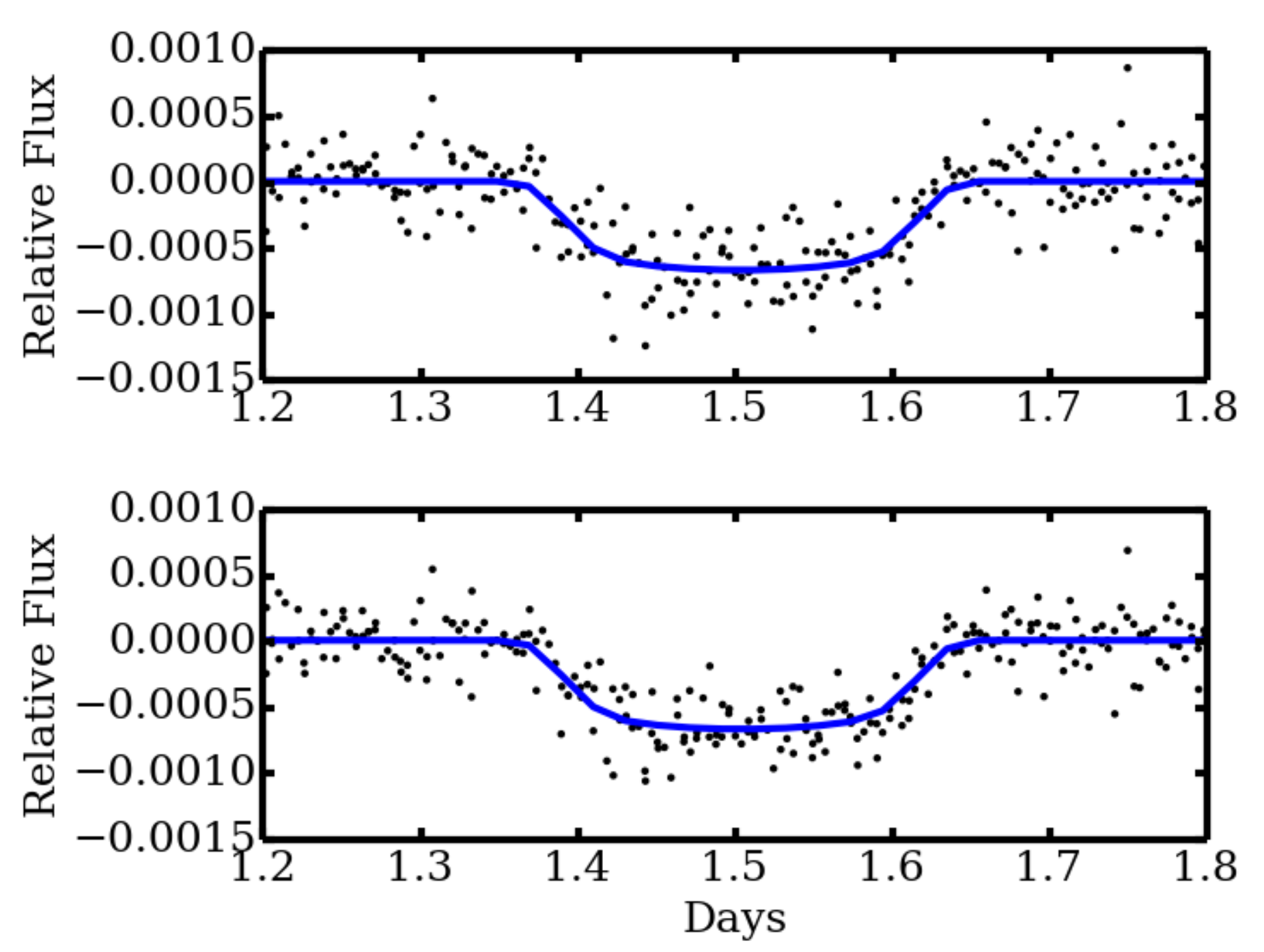}
\caption{Left: Two examples of transits in the EPIC 211351816 lightcurve. Detrended K2 observations of \thestar{} are shown as black dots. The best fit transit model has been plotted in red. The best-fit Gaussian process estimation to the residual lightcurve with transits subtracted is shown in green. The best-fit combined transit + GP model is shown in blue, with 1 and 2 $\sigma$ errors given by the blue contours. The calculation of the relevant values is described in Section 4.1. Top Right: The lightcurve folded at the orbital period of the planet. The best fit transit model has been overplotted in dark blue. Bottom right: The lightcurve folded at the orbital period of the planet, after the best-fit GP model has been subtracted. The decrease in scatter is clearly visible.
\label{GPandtransit}}
\end{figure*}

Equations (1)--(4) are not exact, particularly for stars that are significantly more evolved than the Sun. Empirical tests using interferometry and open clusters and individual frequency modeling have illustrated that the relations typically hold to $\sim$\,5\% in radius and $\sim$\,10\% in mass. Comparisons to model frequencies have also demonstrated that the \dnu\ scaling relation shows systematic deviations of up to a few percent as a function of \teff\ and \feh{} \citep{white11}. We accounted for this through the correction factor $f_{\Delta \nu}$ in Equations (1)--(4), which we determined by iterating the spectroscopic \teff\ and \feh{}\ as well as the asteroseismic mass and \logg\ using the model grid by \citet{sharma16}. The converged correction factor was $f_{\Delta \nu}=0.994$, and our final adopted values for the stellar radius, mass, \logg\ and density are listed in Table \ref{tbl-star}. 

To estimate a stellar age, which cannot be derived from scaling relations alone, we used evolutionary tracks from \citet{bressan12}. Matching the asteroseismic radius to an isochrone with the best-fit asteroseismic mass and \feh{} =\,\fehval{} dex from spectroscopy (see Table \ref{tbl-2}) yielded $\sim$7.8 $\pm$ 2 Gyr. An independent analysis using the BAyesian STellar Algorithm (BASTA), which is based a grid of BaSTI models and has been applied to model several dozen Kepler exoplanet host stars \citep{silvaaguirre2015}, yielded strongly consistent results. The stellar age can be constrained more precisely by modeling individual asteroseismic frequencies, but such modeling is beyond the scope of this paper.

A model-independent estimate of the distance was found using the bolometric flux of 3.579 $\pm$ 0.086 $\times$ 10$^{-13}$ W m$^{-2}$ (uncorrected for extinction) computed from the flux-calibrated spectrum ($\S$ 2.3), the temperature from the high-resolution spectroscopic analysis ($\S$ 3.1), a reddening value of $E(B-V)$ = 0.039 based on the maps of \citet{schlafly2011} and the extinction law of \citet{fitzpatrick1999}.  The estimated distance is 763 $\pm$ 42 pc, placing the star 350 pc above the galactic plane ($b = 27\deg$).  The location well above the plane is consistent with the locations of other RGB stars \citep{casagrande2016} and justifies our use of the $\infty$ value for reddening.
















\section{Lightcurve Analysis and Planetary Parameters}

\subsection{Gaussian process transit model}


The transit of \thestar{b} was first identified by applying the box least-squares algorithm of \citet{kovacs2002} to all targets in our K2 Campaign 5 program. The transits are sufficiently deep to be spotted by eye (see Figure \ref{lightcurve}) and the combined signal to noise is greater than 20, well above commonly adopted thresholds for significant transit events. The transit event was also identified in the planet candidate paper of \citet{pope2016}.



Evolved stars show correlated stellar noise on timescales of hours to weeks due to stellar granulation \citep{mathur2012}, leading to significant biases in transit parameter estimation \citep{carter2009, barclay2015}. To account for this, we used Gaussian process estimation, which has been successfully applied to remove correlated noise in transmission spectroscopy, Kepler lightcurves, and radial velocity data in the past \citep{gibson2012,dawson2014,haywood2014,barclay2015,grunblatt2015}. This is accomplished by describing the covariance of the time-series data as an N$\times$N matrix $\mathbf{\Sigma}$ where 
\begin{equation}
\Sigma_{ij} = \sigma_i^2\delta_{ij} + k(t_i,t_j) 
\end{equation}
where $\sigma_i$ is the observational uncertainty, $\delta_{ij}$ is the Kronecker delta, and $k (t_i, t_j)$ is the so-called covariance kernel function that quantifies the correlations between data points. The simplest and most commonly used kernel function, the squared-exponential or radial basis function kernel, can be expressed as
\begin{equation}
k(t_i,t_j) = h^2 \mathrm{exp}\bigg[ - \Big(\frac{{t_{i} - t_{j}}}{\lambda}\Big)^2\bigg]
\end{equation}
where the covariance amplitude $h$ is measured in flux units
and the length scale $\lambda$ is measured in days \citep{rasmussen2006}. Previous transit studies have used the squared exponential kernel to remove correlated noise without removing the transit signal \citep{barclay2015}.



To analyze the lightcurves, initial parameter guesses are selected for the kernel function, and then a likelihood of the residuals defined by the kernel function parameters is calculated, where the residuals are equivalent to the lightcurve with a Mandel-Agol transit model subtracted from it \citep{mandel2002}. The logarithm of the posterior likelihood of our model is given as
\begin{equation} \label{eq:2}
\mathrm{log}[\mathcal{L}(\mathbf{r})] = -\frac{1}{2}\mathbf{r}^\mathrm{T}\mathbf{\Sigma}^{-1}\mathbf{r} - \frac{1}{2} \mathrm{log} |\mathbf{\Sigma}| - \frac{n}{2} \mathrm{log}(2\pi),
\end{equation}
where $\mathbf{r}$ is the vector of residuals of the data after removal of the mean function (in our case, $\mathbf{r}$ is the lightcurve signal minus the transit model), and $n$ the number of data points. 



\begin{figure*}[ht!]
\epsscale{1.0}
\plotone{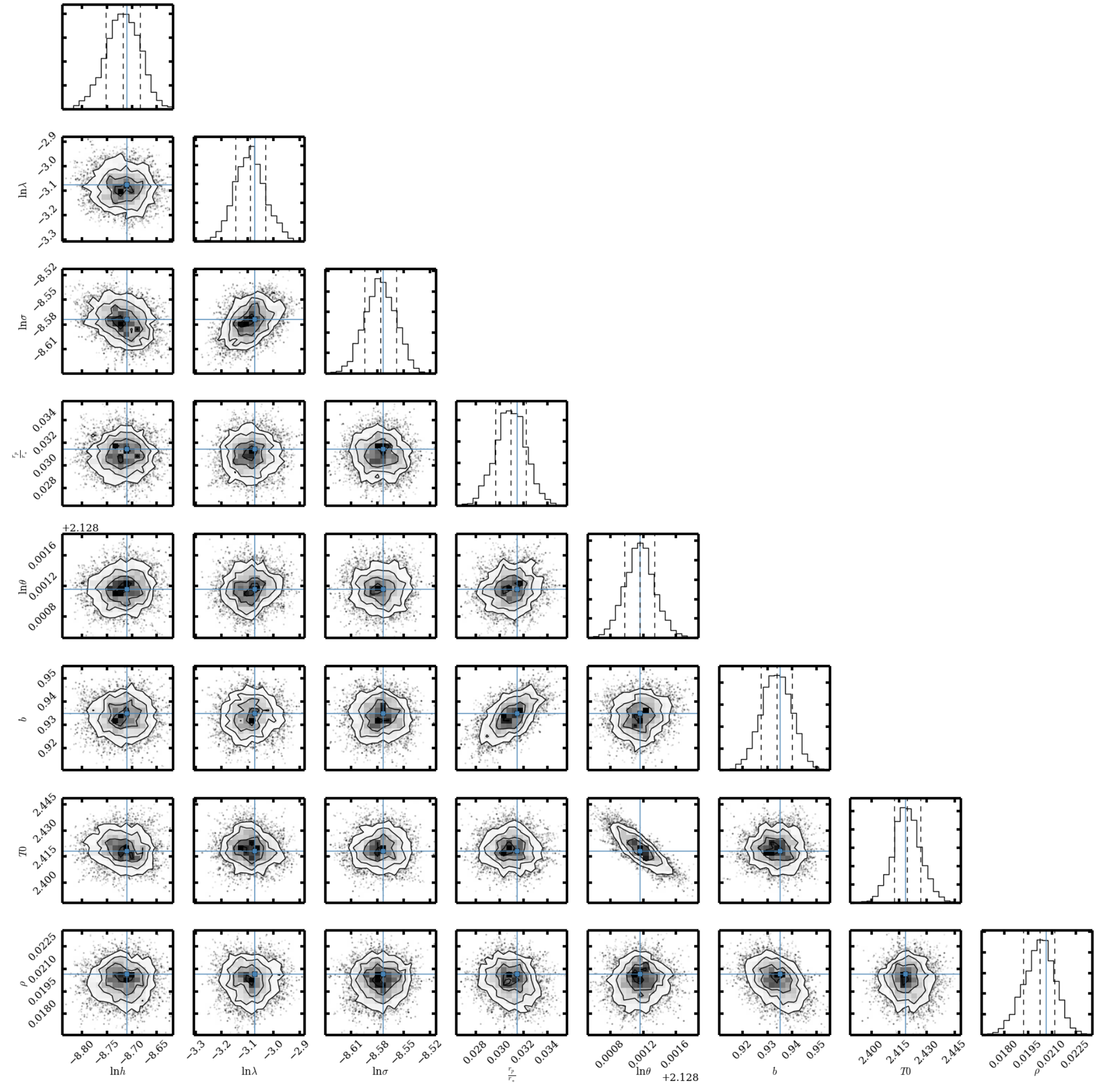}
\caption{Posterior distributions and correlations between all pairs of parameters in our lightcurve MCMC model. Parameters include transit model parameters, squared exponential Gaussian process kernel parameters, and a stellar jitter term. Posterior distributions for each individual parameter are given along the diagonal. 2D contour plots show the correlations between individual parameter pairs. Blue lines correspond to median values. Dotted lines correspond to mean values and standard deviations from the mean. We find that our estimation of the transit depth is not strongly correlated with the other parameters in our model.
\label{triangle}}
\end{figure*}

\begin{figure}[ht!]
\epsscale{1.25}
\plotone{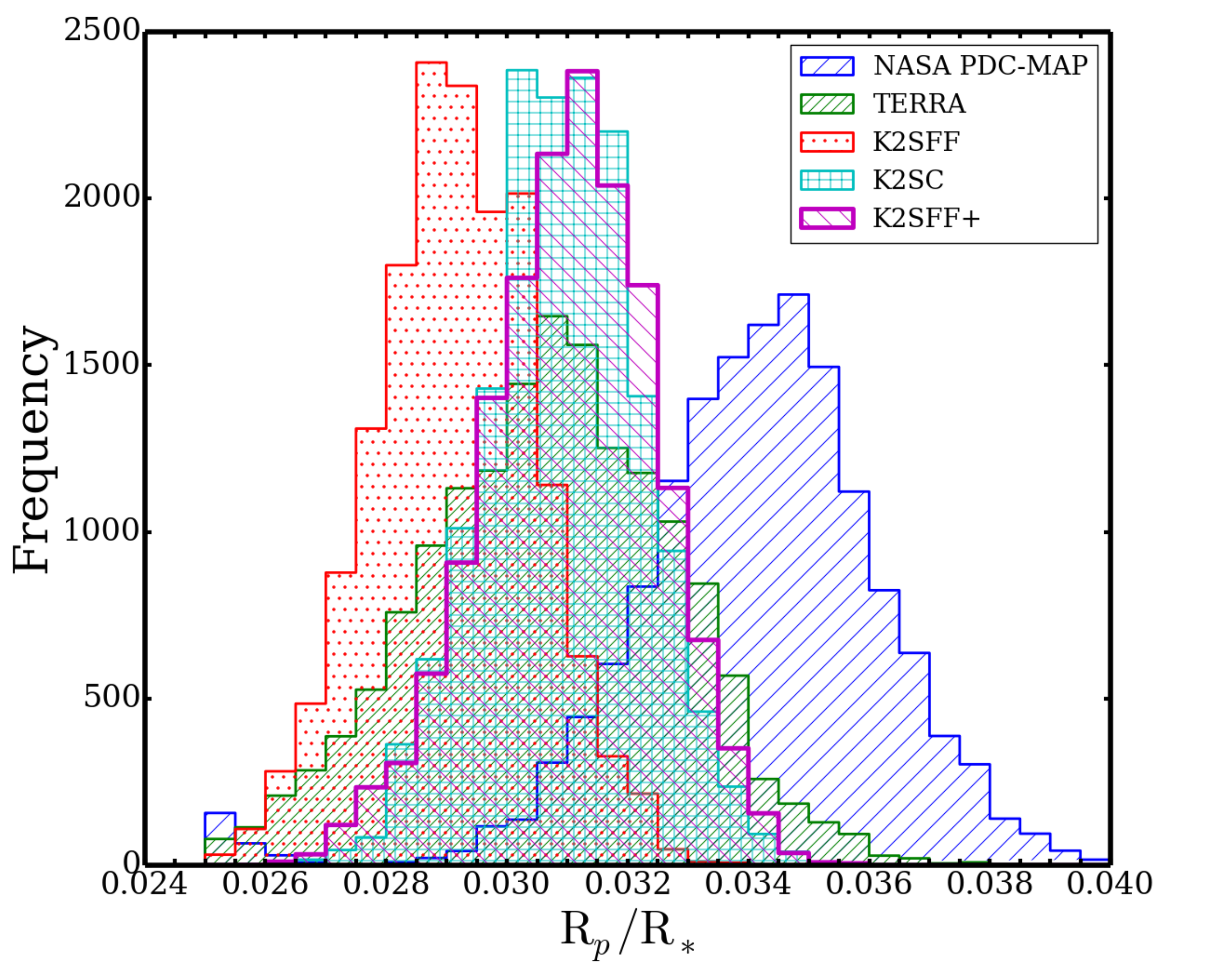}
\caption{Recovered star-to-planet ratios for the \thestar{b} event using lightcurves produced with five different detrending algorithms. We find that the K2SFF lightcurve created with the algorithm of \citet{vanderburg2016} produces the smallest planet to star ratios on average, while the NASA PDC-MAP lightcurve produces a planet to star ratio considerably larger than the other detrending algorithms. We choose the lightcurve where transits and instrumental effects were fit simultaneously for subsequent analysis, as a transit injection/recovery test comparing this K2SFF+ method and the standard K2SFF method revealed that transit depths were diluted by the standard K2SFF detrending but retained by the simultaneous K2SFF detrending and transit fit method. \label{hist}}
\end{figure}

 The GP kernel function and transit model parameters are then fit as free parameters via Markov chain Monte Carlo (MCMC) exploration of parameter space using the Python software package \texttt{emcee} \citep{foremanmackey2013}. The \texttt{emcee} package contains an Affine-invariant MCMC Ensemble sampler, which determines the maximum likelihood parameters  through an iterative exploration of parameter space. We draw the planet radius from this MCMC exploration of parameter space, with 1-$\sigma$ error corresponding to 68$\%$ confidence intervals in the MCMC distributions of all free parameters. Along with the planet-to-star radius ratio, the impact parameter, period, and ephemeris of transit were fit simultaneously with the Gaussian process kernel parameters and a photometric jitter term. Limb darkening parameters were fixed to the \citet{claret2011} stellar atmosphere model grid values closest to the measured temperature, surface gravity, and metallicity of the host star. Initial parameter values and priors were determined via a least squares transit fit using \texttt{ktransit} \citep{TomBarclay2015}. The results and priors for this simultaneous parameter fitting are listed in Table \ref{tbl-2} and parameter distributions are given in Figure \ref{triangle}. 
 
 
To ensure our results were replicable, we performed a second MCMC analysis of the system using additional model parameters using a method very similar to that applied to Kepler-91 by \citet{barclay2015}. Mean stellar density, photometric zeropoint, two limb darkening parameters, radial velocity zero point, two Gaussian process hyperparameters, time of mid-transit, orbital period, impact parameter, the scaled planet radius, two eccentricity vectors ($e\sin{\omega}$ and $e\cos{\omega}$), radial velocity semi-amplitude, secondary eclipse depth, amplitude of ellipsoidal variations, amplitude of reflected light from the planet, and two uncertainty parameters added in quadrature with the reported uncertainties on radial velocity and photometric data were included in this secondary model. The priors on these parameters were uniform except for a Gaussian prior based on the asteroseismic value of the mean stellar density, priors that kept the two limb darkening parameters physical \citep{burke2008} plus Gaussian priors with means taken from \citet{claret2011} and a standard deviation of 0.4,  a prior of $1/e$ on the eccentricity to avoid biasing this value high \citep{Eastman2012} and an additional prior that took the form of a Beta function with parameters determined by \citet{vaneylen2015}. Additionally, we sampled the logarithm of the Gaussian process hyperparameters, RV semi-amplitude, secondary eclipse depth, ellipsoidal variations, reflected light, and two uncertainty parameters. We ran the MCMC algorithm using 600 walkers and 20,000 steps yielding 12 million samples. We found posteriors on the scaled planet radius of $0.0296^{+0.0035}_{-0.0024}$ and an impact parameter of 0.921$^{+0.023}_{-0.032}$, strongly consistent with our earlier study. A secondary eclipse, ellipsoidal variations and any reflected light from the planet were not detected. We found an eccentricity of a few percent, marginally inconsistent with zero.


 \begin{deluxetable*}{ccrrrrrrrrcrl}
\tabletypesize{\scriptsize}
\tablecaption{Posterior Probabilities from Lightcurve and Radial Velocity MCMC Modeling\label{tbl-2}}
\tablewidth{0pt}
\tablehead{
\colhead{Parameter} & \colhead{Median} & \colhead{84.1$\%$} & \colhead{15.9 $\%$} & \colhead{Prior}
}
\startdata
$\rho$ (g cm$^{-3}$) & 0.020 & +0.001 & -0.001 & $\mathcal{N}$(0.02; 0.001) \\
T$_0$ (BKJD) & 2309.072 & +0.007 & -0.007 & $\mathcal{U}$(1.3; 2.5) \\
$P_{\mathrm{orb}}$ (days) & 8.4062 & +0.0015 & -0.0015 & $\mathcal{U}$(8.3; 8.5) \\ 
$b$ & 0.933 & +0.006 & -0.007 & $\mathcal{U}$(0.0, 1.0 + $R_p/R_*$)\\ 
$R_p / R_{*}$ & 0.0311 & +0.0013 & -0.0015 & $\mathcal{U}$(0.0, 0.5) \\
K (m s$^{-1}$) & 103 & +8 & -8 & \\
T$_{0,\mathrm{RV}}$ (BKJD) & 2583.808 & +0.007 & -0.007 & $\mathcal{U}$(0.0, $P_{\mathrm{orb}}$) \\
ln$f$ & -3.8 & +2.8 & -3.9 & $\mathcal{U}$(-10, 10)\\
$h_{\mathrm{GP}}$ (ppm) & 157 & +5 & -5 & $\mathcal{U}$(exp(-12, 0))\\
$\lambda_{\mathrm{GP}}$ (days) & 0.057 & +0.005 & -0.004 & $\mathcal{U}$(exp(-10, 10))\\
 $\sigma_{\mathrm{GP}}$ (ppm) & 189 & +4 & -4 &  $\mathcal{U}$(exp(-20, 0))\\
\enddata 
\tablecomments{$\mathcal{N}$ indicates a normal distribution with mean and standard deviation given respectively. $\mathcal{U}$ indicates a uniform distribution between the two given boundaries. Ephemerides were fit relative to the first measurement in the sample and then later converted to Barycentric \emph{Kepler} Julian Date (BKJD). Transit limb darkening parameters $\gamma_1$ and $\gamma_2$ were fixed to 0.6505 and 0.1041, respectively.}

\end{deluxetable*}

\subsection{Radial Velocity Analysis: Planetary Confirmation and False Positive Assessment}

\begin{figure}[ht!]
\epsscale{1.2}
\plotone{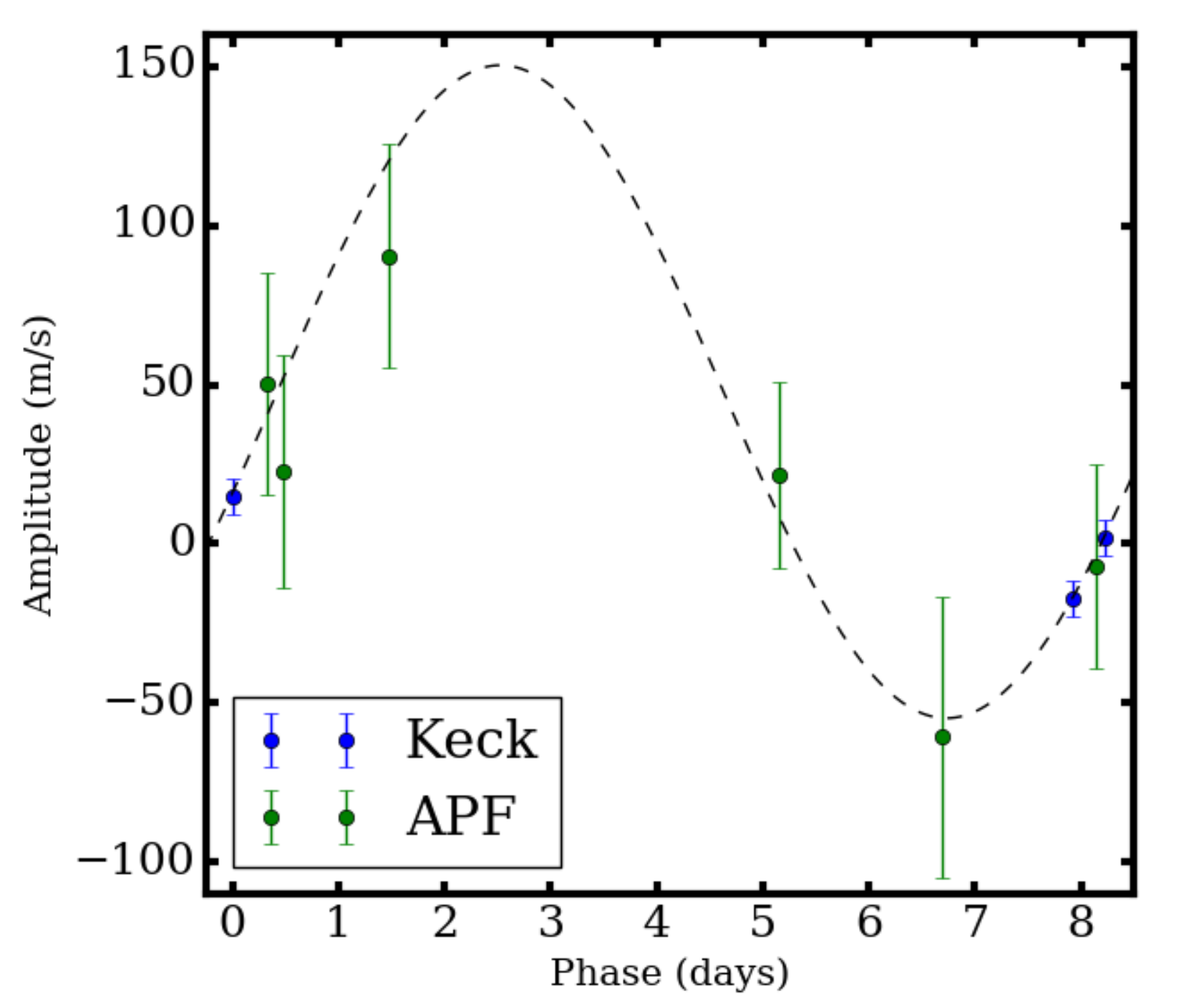}
\caption{Radial velocity measurements of the system, phase-folded at the known orbital period. The initial measurements obtained with Keck/HIRES are shown in blue and have errors which are smaller than the markers in the plot. The remaining green measurements were taken with the Levy spectrometer on the Automated Planet Finder telescope. The dashed gray curve corresponds to a one-planet Keplerian orbit fit to the data. The best fit Keplerian orbital parameters were found using $\texttt{emcee}$. A stellar jitter term of 5 m s$^{-1}$ was added in quadrature to make measurement errors more robust.
\label{rv}}
\end{figure}

We modeled the APF and Keck radial velocity measurements of the planet with a Keplerian orbital model. Assuming \thestar{b} would produce the dominant signal in the radial velocity measurements, we assume a circular orbit for the planet and fit the data with a sinusoid with a period set to the orbital period obtained from the transit fitting. Using a Markov chain Monte Carlo method, best-fit values were determined for the phase and amplitude of the radial velocity variations. We applied a velocity shift of 23 m s$^{-1}$ to the APF measurements relative to the Keck measurements, and additionally fit for a non-zero offset to the resultant sinusoid to account for the different RV zero points of the two instruments. The mass of the planet was then estimated from the Doppler amplitude. The best fit RV model and relative measurement values are shown in Figure \ref{rv}. As subgiant and giant stars are known to have an additional 4-6 m s$^{-1}$ of velocity scatter due to stellar jitter \citep{johnson2007a}, we adopted a value of 5 m s$^{-1}$ and add it to our measurement errors in quadrature.

The \emph{Kepler} pixels span 4" on the sky, and thus background eclipsing binaries (EBs) can often cause false positive transit signals \citep{jenkins2010, batalha2010, everett2015}. In addition, the K2 lightcurve was constructed using an aperture that is 7 pixels or 28'' across, exacerbating the possibility of a false positive. As the maximum transit depth of an EB is 50\%, such a system would have to be at least as bright as Kepler magnitude ($K_P) \approx 19$ to mimic a transit.  To identify potential culprits, we searched the photometry database of the Sloan Digital Sky Survey (Data Release 9) for sources within 30'' of \thestar{}.  We identified only a single source (SDSS J083104.13+105112.9) of interest.  It has an estimated $K_P = 19.05$, yet is well outside the photometric aperture and the small fraction of light scattered into the aperture by the {\it Kepler} point response function ensures it could not have produced the transit signal.  No sources were detected in our Keck 2-NIRC 2 AO imaging down to $K'=15.5-18$ (0.2-2''), corresponding to $K_P > 19$ for M dwarf stars that are the most likely components of faint background EBs. 

To calculate a false positive probability for the background EB scenario, we followed the method of \citet{Gaidos2016}.  This discrete (Monte Carlo) Bayesian calculation uses a synthetic population generated by the TRILEGAL galactic stellar population model as priors \citep[v. 1.6;][]{Vanhollebeke2009} for 10 square degrees at the location of \thestar{} on the sky.  Likelihoods are calculated by imposing constraints on stellar density from the transit duration and orbital period, and on brightness from the non-detections in the SDSS and NIRC2 images, requiring that the diluted eclipse depth is at least equal to the transit depth.  We found that the false positive probability for this scenario is effectively zero, as no star from the simulated background population can simultaneously satisfy the stellar magnitude and density constraints.  Background stars are either too faint to produce the transit or are ruled out by our high-resolution imaging, and the long transit duration implies a stellar density that is too low for dwarf stars\footnote{Long transit durations can occur at the apoapsis of highly eccentric orbits, but such orbits would have been circularized by the $\sim$7 Gyr age of this system.}. Low stellar density precludes a companion dwarf EB as the source of the signal; evolved companions are ruled out by our AO imaging to within 0.2'' and stellar counterparts within $\sim 1$ AU are ruled out by the absence of a drift in our radial velocity data.   

\section{Discussion}


\subsection{Is EPIC 211351816.01 Inflated?}

We have described the discovery and characterization of a Jupiter-mass planet on an 8.4-day orbit around a red giant branch star. This object joins a sample of only five other known transiting planets hosted by highly evolved stars \citep{huber2013b, lillo-box2014, barclay2015, quinn2015, ciceri2015, vaneylen2016}. The high metallicity of the host star is also characteristic of the close-in gas giant planet population, suggesting that this system may be simply a successor to such ``hot Jupiter" systems.  

As the stellar radius of \thestar{} has been determined to 3\% precision through asteroseismology, the dominant uncertainty in planet radius for this system comes from the transit depth. We compared the star-to-planet radius ratio ($R_p /R_*$) for this system using lightcurves produced by the PDC-MAP pipeline \citep{stumpe2012,smith2012}, the K2 ``self flat field" (K2SFF) pipeline \citep{vanderburg2016} as well as a modified version of the \citet{vanderburg2016} pipeline which simultaneously fit thruster systematics, low frequency variability, and planet transits with a Levenberg-Marquardt minimization algorithm, the K2SC pipeline \citep{aigrain2016}, and the TERRA pipeline \citep{petigura2013}. We find that measured transit depths varies by over 30\% between the different systematic detrending pipelines we tested. We plot the spread in recovered star-to-planet radius ratios in Figure \ref{hist}. 

To investigate the differences in $R_p / R_*$ recovered from lightcurves produced from different pipelines, we injected transits modeled from those in the \thestar{} system into lightcurves (with systematics) of 50 stars classified as low-luminosity red giants from our K2 Campaign 5 target list. These lightcurves were then detrended using both the standard K2SFF method of \citet{vanderburg2016} as well as the modified method which detrended instrumental noise and fit the planet transit simultaneously (hereby referred to as K2SFF+). The transit depths in both sets of processed lightcurves were then fit using a box least squares search \citep{kovacs2002} and a Mandel-Agol transit model \citep{mandel2002, TomBarclay2015}. This transit injection/recovery test revealed that the transit depth was retained with some scatter when both the transit and systematics were fit simultaneously, but when the systematics were fit and removed with the nominal \citet{vanderburg2016} method, transit depths were reduced by 13\% and the planet's radius was underestimated by 8\% on average. 

We report results from the K2SFF+ lightcurve as it was demonstrated to preserve transit depth through our transit injection/recovery tests, and its measured transit depth is strongly consistent with transit depths measured from two independently detrended lightcurves. We add an additional 5\% error in planet radius to account for the uncertainty in transit fitting seen in the injection/recovery tests. Current and future studies with injection/recovery tests similar to those performed for Kepler \citep{petigura2013, christiansen2015} will help resolve this discrepancy between accuracy and precision in measuring transit depths with K2.



\subsection{Planet Inflation Scenarios}

 \begin{figure*}[ht!]
\epsscale{1.1}
\plottwo{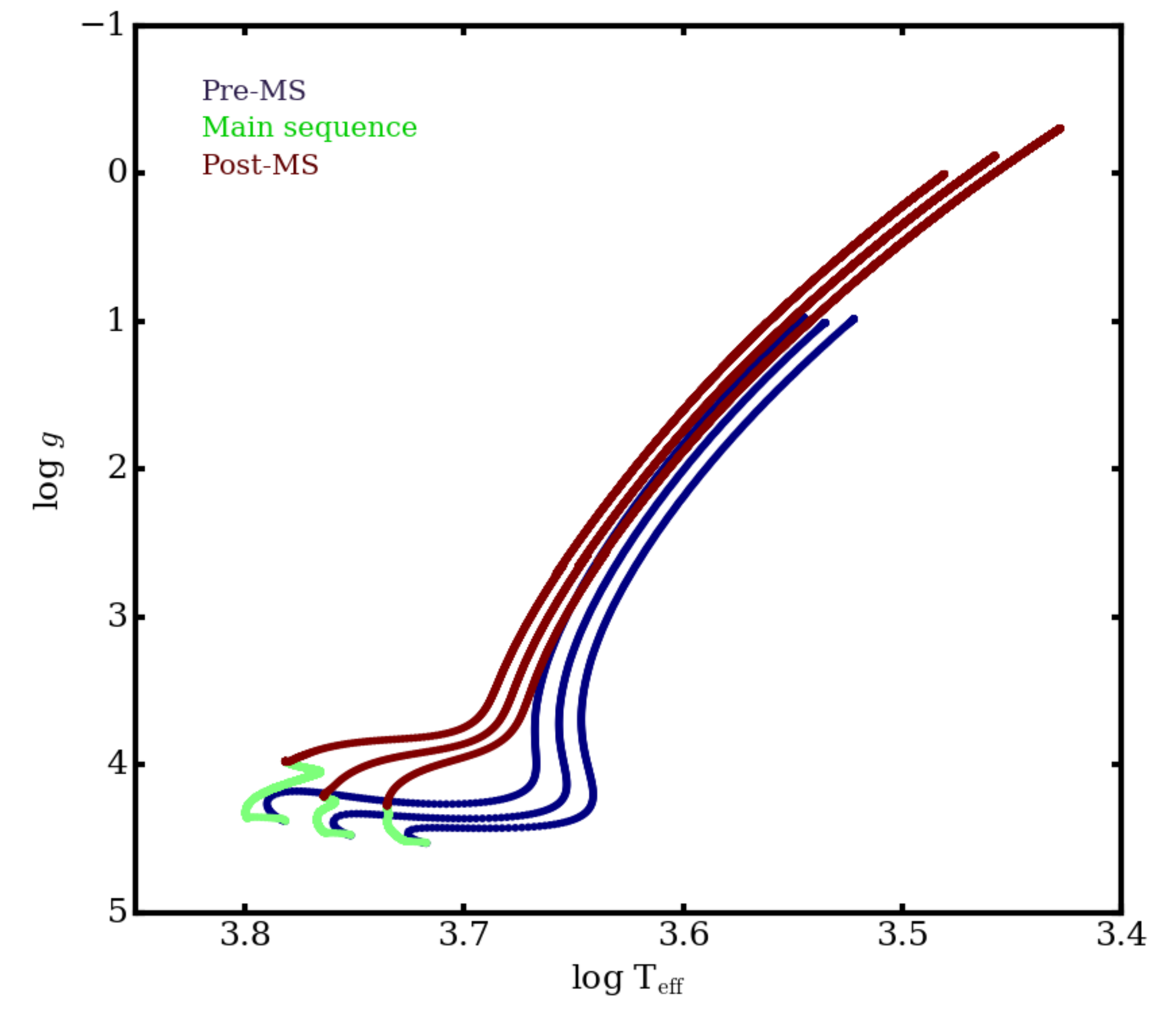}{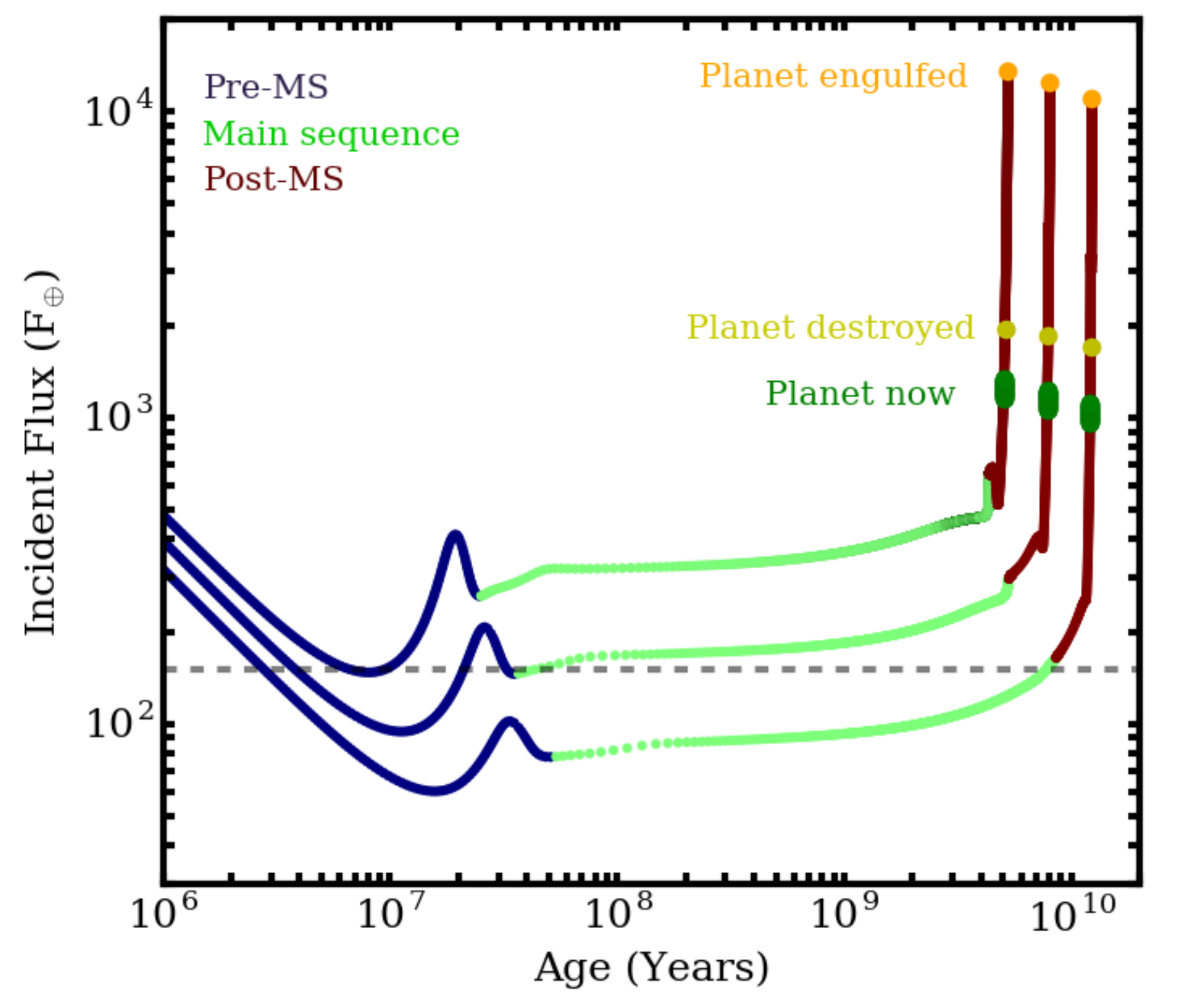}
\caption{Left: Surface gravity versus effective temperature for 1.0 (rightmost), 1.15, and 1.3 M$_\odot$ (leftmost) Parsec evolutionary tracks with $\rm{[Fe/H]}$ = 0.60, \fehval, and 0.34 dex, respectively. Note that the choice of mass and metallicity correspond to lower and upper bounds for the stellar characteristics of \thestar{}. Blue, green, and red correspond to pre-main sequence, main sequence, and red giant branch stages of stellar evolution. Right: Change in incident flux on \thestar{b} over time for the models shown in the left panel.  The current incident flux on the planet, assuming a stellar radius constrained by asteroseismic measurement, is denoted by dark green. The point at which the planet will be engulfed is denoted in orange, and tidally disrupted noted in yellow (see \S 5.3). The gray dotted line corresponds to the inflation threshold as cited by \citet{lopez2016}. \label{evol}}
\end{figure*}

\begin{figure}[ht!]
\epsscale{1.25}
\plotone{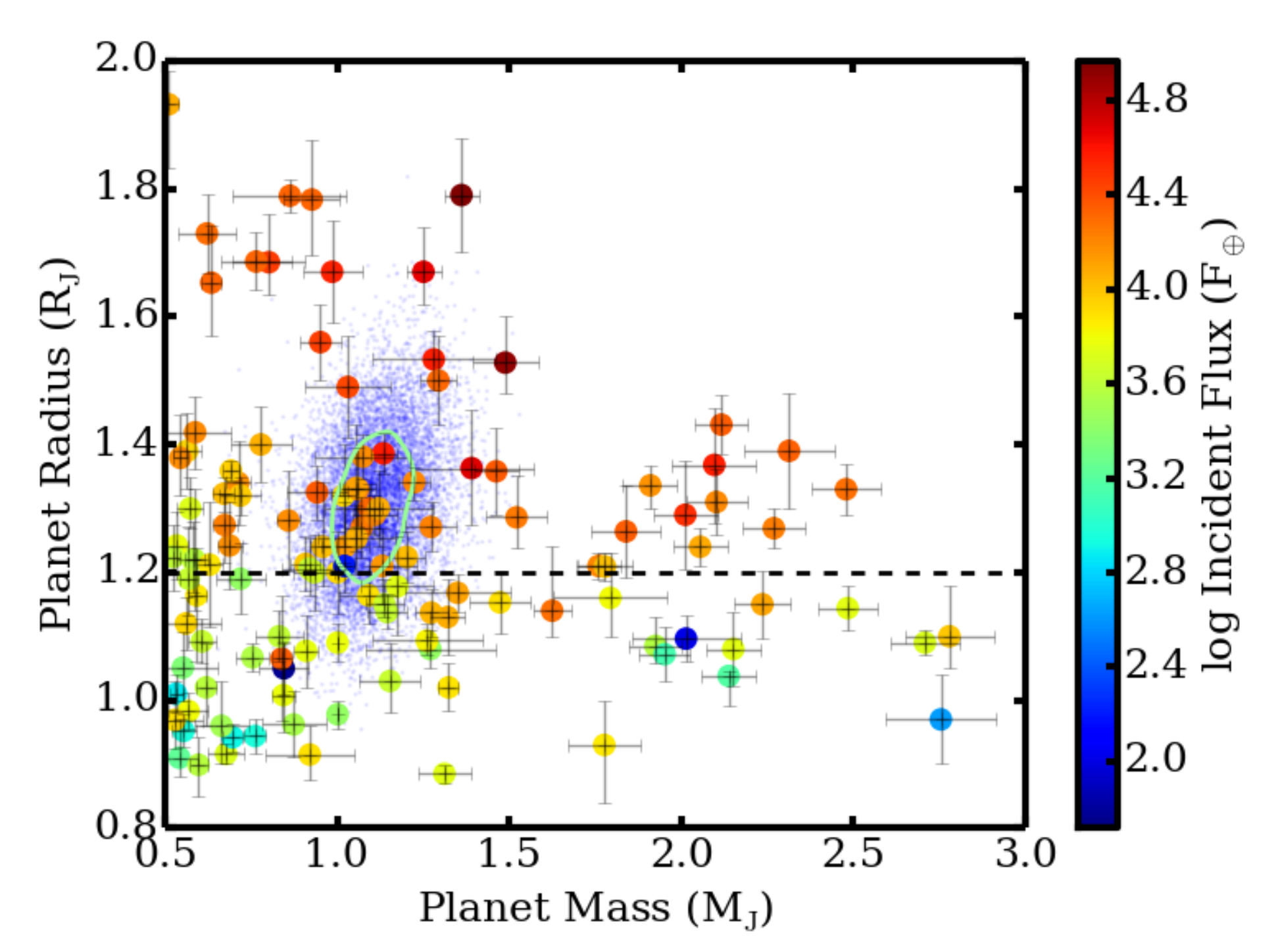}
\caption{Planet mass versus radius in units of Jupiter mass and radius for well characterized planets with errors of less than 0.1 Jupiter radii and 0.2 Jupiter masses. The dotted line shows the approximate threshold of planet inflation, as given by \citet{lopez2016}. Color shows the logarithm of the incident flux in units of Earth fluxes. \thestar{b} is shown as the cloud of points near 1.25 R$_{\mathrm{J}}$ and 1.1 M$_{\mathrm{J}}$, with 1-$\sigma$ errors shown by the teal contour. The color of points in the cloud correspond to the incident flux \thestar{b} received on the main sequence, which is clearly uncharacteristic of the known, well-characterized inflated planets, suggestive of a non-inflated past. The color of the contour indicates its current incident flux. Planet characteristics have been taken from the Exoplanet Orbit Database and the Exoplanet Data Explorer at exoplanets.org. 
\label{mrdiag}}
\end{figure}

\begin{figure}[ht!]
\epsscale{1.25}
\plotone{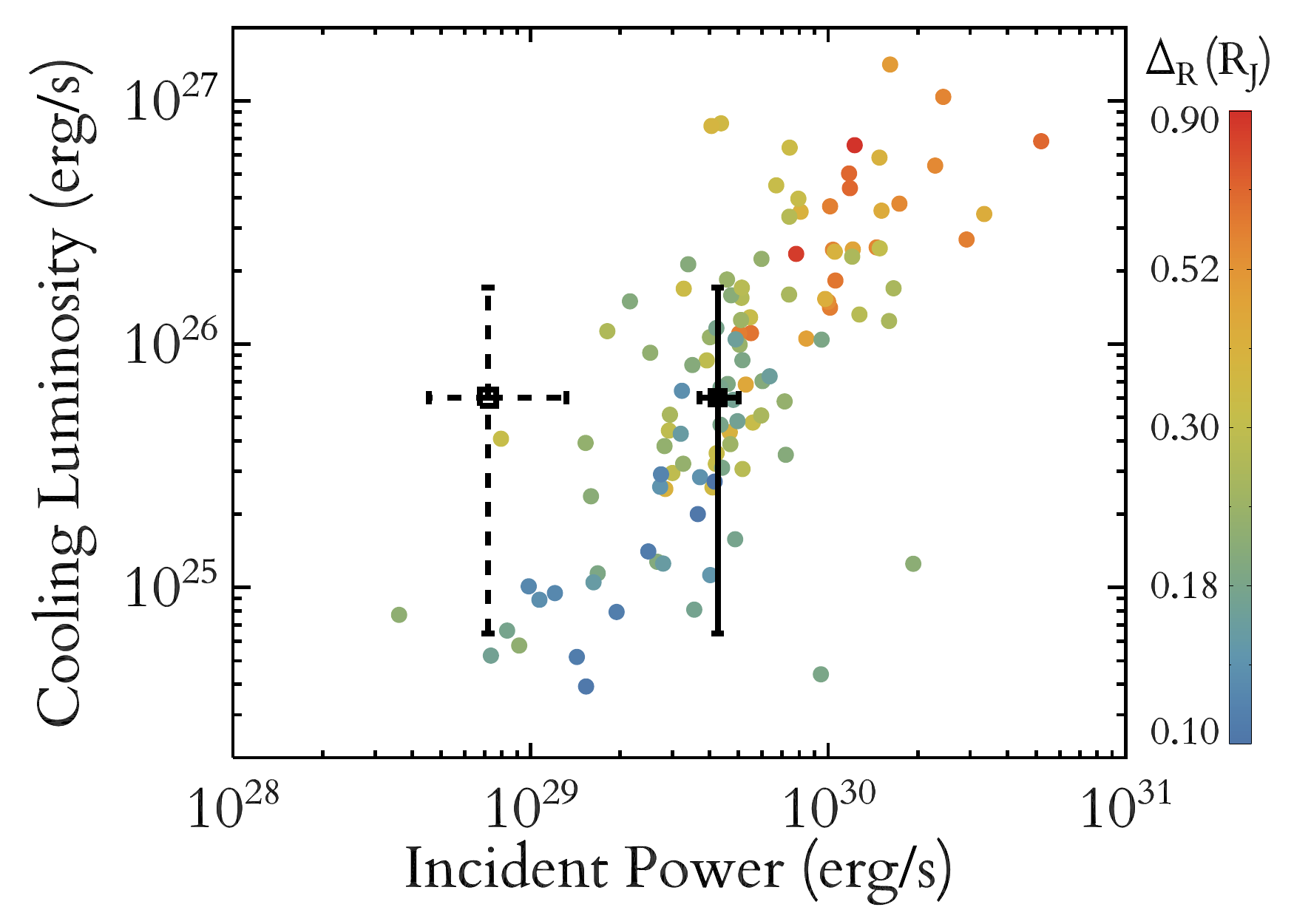}
\caption{Steady-state cooling luminosity, or the power the planet must emit to retain its measured radius, as a function of incident power, with radius anomaly, or the difference in radius between measured and predicted planet size indicated in color. Predicted planet sizes have been calculated assuming a planet of pure H/He using the models of \citet{lopez2016}. The filled square with solid error bars shows \thestar{b} at its current incident flux, whereas the open square with dashed error bars show the planet at its main sequence incident flux.  The current cooling luminosity of the planet is characteristic of the inflated planet population around main sequence stars, suggesting that the physical mechanism inflating this planet is the same. However, the planet would be inflated to an uncharacteristically high degree if it were to maintain its current radius around a main sequence star. The planet seen nearest to this case on the plot is WASP-67b, a young, 0.47 M$_{\mathrm{J}}$ planet, whose significantly lower mass allows it to be more easily inflated. Inflating the more massive \thestar{b} to the same degree as WASP-67b should require an incident power higher than the \thestar{b} receives now. \label{inflatedtcool}}
\end{figure}


\begin{figure}[ht!]
\epsscale{1.2}
\plotone{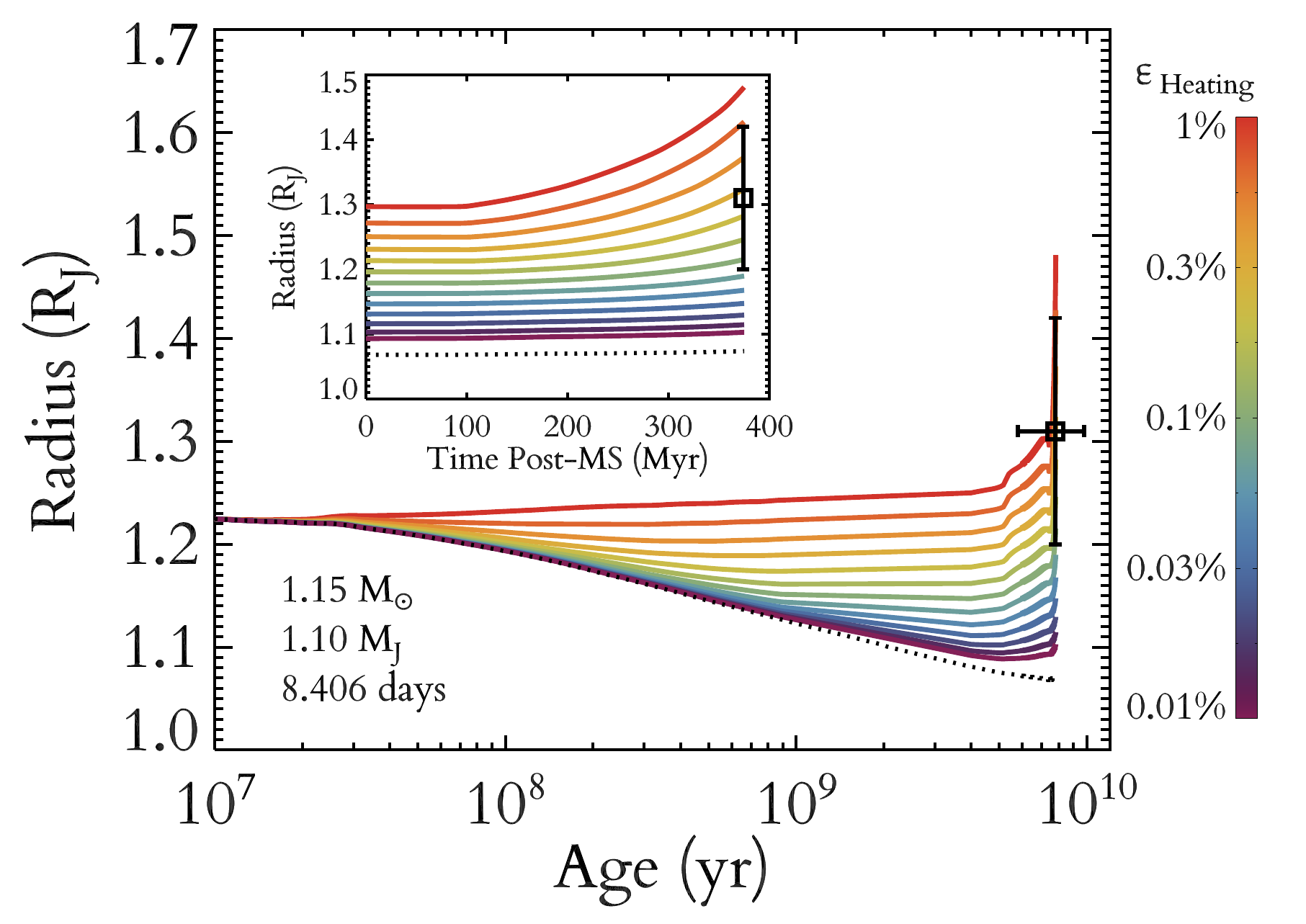}
\caption{Planetary radius as a function of time, shown for various potential heating efficiencies. We assume the best-fit values for the stellar mass and the planetary mass and radius, and a planetary composition of a H/He envelope surrounding a 20 M$_\oplus$ core of heavier elements. The dotted line corresponds to a scenario with no planetary heating. The inset shows the post-main sequence evolution at a finer time resolution. The measured planet radius is consistent with heating efficiencies of 0.1 to 0.5$\%$, and inconsistent with the class II, delayed cooling scenario. 
\label{heating}}
\end{figure}

We can test planet inflation mechanisms by examining the response of planets to increasing irradiation as the host star leaves the main sequence. In particular, planets with orbital periods of $<$30 days will experience levels of irradiation comparable to typical hot Jupiters for more than 100 Myr. Following the nomenclature of \citet{lopez2016}, if the inflation mechanism requires direct heating and thus falls into Class I, the planet's radius should enter a re-inflated state around a post-main sequence star. However, if the inflation mechanism falls into Class II, requiring delayed cooling, there should be no effect on planet radius as a star enters the red giant phase, and re-inflation will not occur. \thestar{b} provides a valuable test for the re-inflation hypothesis, as it is inflated now but orbits at a distance such that it may not have received irradiation above the inflation threshold for its entire existence. 

To estimate the change in stellar irradiation over time, we use the Parsec evolutionary tracks \citep{bressan12} with the host star mass and metallicity derived in \S 3.2. Figure \ref{evol} shows an HR diagram and incident flux evolution for models with masses of 1.0, 1.15 and 1.3 M$_\odot$ from the pre-main sequence to the tip of the red giant branch. We used metallicities of 0.6, 0.42 and 0.34 dex for the 1.0, 1.15 and 1.3 M$_\odot$ models, respectively, which results in overestimated limits given that metal-poor stars are hotter than metal-rich stars for a fixed mass. We also denote an inflation threshold of 2 $\times$ 10$^8$ erg s$^{-1}$ cm$^{-2}$ ($\sim$150 F$_\oplus$) following \citet{demory2011} and \citet{miller2011}, who note that this corresponds to an equilibrium temperature of 990 K assuming a Bond albedo of 0.1, comparable to the temperature at which Ohmic heating may become important \citep{batygin2011}. None of the 38 transiting giant planets with insolations below this threshold known to date appear to be inflated \citep{thorngren2015}. 


Figure \ref{evol} demonstrates that the incident flux of this planet may have been above the 150 F$_\oplus$ threshold for inflation throughout its main sequence life. However, it is also possible that the planet experienced a flux below this threshold, depending on the exact mass and metallicity of the star. To estimate the main-sequence incident flux level quantitatively, we performed Monte Carlo simulations by interpolating the evolutionary tracks to randomly sampled values of stellar mass and metallicity as measured for \thestar{} and calculated the average incident flux on the main sequence. The resulting distribution yielded an average main sequence flux of 170$^{+140}_{-60}$ F$_\oplus$. We also estimated the incident flux evolution using a different set of evolutionary tracks from the MIST database \citep{choi2016}, which yielded consistent results. Our analysis demonstrates that EPIC 211351816.01 received a main-sequence incident flux which is close to the inflation threshold, but lower than the typical incident flux for planets with a comparable radius. This suggests that additional inflation occurred after the star evolved off the main sequence.



We illustrate the current constraints on the mass and radius of \thestar{b} in Figure \ref{mrdiag} relative to other known, well-characterized giant planets. The dotted line denotes the empirical threshold for planet inflation put forth by \citet{miller2011}. Colors correspond to the incident fluxes on these planets, except in the case of \thestar{b} where we have also indicated the incident flux the planet would have received on the main sequence to illustrate how uncharacteristic of the inflated planet population it would have been at that time.

Furthermore, the energetics of \thestar{b} indicate that if it was inflated to its current radius while its host star was on the main sequence, the planet would be an outlier within the inflated planet population, with internal heating over an order of magnitude higher than would be expected. We illustrate this in Figure \ref{inflatedtcool}, where we plot the intrinsic cooling luminosity predicted by the models of \citet{lopez2016} against incident flux for the known inflated planet population. The radius anomaly, or difference in measured and predicted planet size, is indicated by color. The filled square corresponds to \thestar{b} today, showing clear agreement with the rest of the inflated planet population energetically. However, the open square with dashed error bars corresponds to the incident flux on the planet when its host star was on the main sequence. The only planet energetically comparable to this scenario is WASP-67b, a planet with less than half the mass around a young star \citep{hellier2012}. As lower mass planets are easier to inflate, and young planets may still be inflated from their initial formation, it would be very surprising to find a Jupiter-mass, middle-aged planet with similar energetic qualities. This, along with the empirical evidence for the energetic boundary of inflation of 2 $\times$ 10$^8$ erg s$^{-1}$ established by \citet{miller2011}, suggest that \thestar{b} was not inflated when its host star was on the main sequence.


Assuming that the inflation of the planet was due to the deposition of flux into the planet interior, we can use the model of \citet{lopez2016} to estimate the heating efficiency needed to reproduce the current radius of \thestar{b}. Figure \ref{heating} shows the radius evolution of \thestar{b} as a function of age, given a range of heating efficiencies, a planetary structure of a H/He envelope with a 20 M$_\oplus$ core of heavier elements, and a 1.15 Msun, [Fe/H] = + 0.42 dex model for the star. The scenario with no additional interior heating is shown by the dotted line. The planet is consistent with heating efficiencies of $\sim0.3\%$, and inconsistent with a class II scenario with no additional heating at late times. This suggests \thestar{b} may be the first re-inflated planet discovered. 

Further studies of giant planets around evolved stars will be necessary to confirm this hypothesis. Gas planets at a slightly larger orbital period ($\sim$10--30 days) around a similar star would experience fluxes well below the empirical inflation threshold during the main sequence and would thus provide a clearer picture of the inflation mechanism. Although planets inflated by mechanisms more heavily dependent on factors other than incident flux, such as metallicity, have not been observed around main sequence stars, these factors could potentially delay contraction at orbital distances beyond the nominal inflation boundary, and thus we cannot completely rule out the possibility that such effects may also be responsible for the inflation of this planet  \citep{chabrier2007}.


\subsection{Planetary Engulfment}

The expansion of a star in the red giant phase can extend to AU scales, eventually engulfing any short-period planets. We calculate that \thestar{b} will be engulfed when its host star reaches a radius of $\sim$18 R$_{\odot}$. This provides a conservative upper limit for the remaining lifetime of the planet of $\sim$200 Myr. 

The scarcity of short-period planets orbiting giant stars has been suggested to be a result of tidally-driven orbital decay \citep{schlaufman2013}. We can estimate the timescale of orbital decay due to tides following the prescription of \citet{schlaufman2013}:

\begin{multline}
t = 10 ~\mathrm{Gyr} ~ \frac{Q_* / k_*}{10^6}\Bigg(\frac{M_*}{\mathrm{M}_{\odot}}\Bigg)^{1/2}\Bigg(\frac{M_p}{\mathrm{M}_{\mathrm{Jup}}}\Bigg)^{-1} \\
\times \Bigg(\frac{R_*}{\mathrm{R}_{\odot}}\Bigg)^{-5}\Bigg(\frac{a}{0.06 \mathrm{AU}}\Bigg)^{-13/2}
\end{multline}

Here, $Q_*$ is the tidal quality factor of the star, and $k_*$ its tidal Love number. These values are highly uncertain, but making the usual assumption of $Q_*/k_*$ = 10$^6$ \citep{schlaufman2013} the decay time is $\approx$ 60 Myr. If, however, $Q_*/k_*$ = 10$^2$, as \citet{schlaufman2013} suggest may be the case for sub-giant stars, then $t \approx$ 6,000 yr. This indicates that such a low value for $Q_*/k_*$ is implausible. Consequently, the discovery of \thestar{b} along with other planets around evolved such as K2-39b \citep{vaneylen2016} and Kepler-91b \citep{barclay2015} suggests that observation bias may contribute to the relative paucity of planets detected on short-period orbits around giant stars.

\section{Conclusions}


We report the discovery of a transiting planet with R = \planrad{} and M = \planmass{} around the low luminosity giant star \thestar. We use a Gaussian process to estimate the correlated noise in the lightcurve to quantify and remove potential correlations between planetary and noise properties. We also tested five different lightcurves produced by independent systematic detrending methods to account for inconsistencies in the treatment of K2 data and derive an accurate transit depth and planet radius. We performed an iterative spectroscopic and asteroseismic study of the host star EPIC 211351816 to precisely determine its stellar parameters and evolutionary history.

We determine that, assuming a stable planetary orbit for the range of acceptable stellar parameters, \thestar{b} requires approximately 0.3$\%$ of the current incident stellar flux to be deposited into the planet's deep convective interior to explain its radius. The measured planet radius is inconsistent with most inflation scenarios without current heating of the planet's interior. This suggests planet inflation may be a direct response to stellar irradiation rather than an effect of delayed planet cooling after formation, and \thestar{b} is a strong candidate for the first known re-inflated planet.

Further studies of planets around evolved stars are essential to confirm the planet re-inflation hypothesis. Planets may be inflated beyond the nominal inflation regime by methods that are more strongly dependent on other factors, such as atmospheric metallicity, than incident flux. An inflated planet observed around a giant star with an orbital period of $\sim$20 days would have been outside the inflated planet regime when its host star was on the main sequence, and thus finding such a planet could provide more insight into the re-inflation hypothesis. Using a Gaussian process to characterize stellar noise seen in the lightcurve may allow for the discovery of smaller planets than previously possible around giant stars. Other Gaussian process kernels, or fitting additional transit parameters such as limb darkening coefficients, could provide additional insight. Further study on this particular system, such as a more detailed asteroseismic analysis to determine a more precise age, will provide deeper insight into the evolutionary history of this system and the inflation history of hot Jupiters as a whole. This discovery also motivates new theoretical work exploring exactly how different inflationary heating mechanisms respond to post main sequence changes in irradiation.


\acknowledgements{The authors would like to thank Jeffrey C. Smith, Suzanne Aigrain and Travis Berger for helpful discussions. This research was supported by NASA Origins of Solar Systems grant NNX11AC33G to E.G. and by the NASA K2 Guest Observer Award NNX16AH45G to D.H..  D.H. acknowledges support by the Australian Research Council's Discovery Projects funding scheme (project number DE140101364) and support by the National Aeronautics and Space Administration under Grant NNX14AB92G issued through the Kepler Participating Scientist Program. This research has made use of the Exoplanet Orbit Database and the Exoplanet Data Explorer at Exoplanets.org. This work was based on observations at the W. M. Keck Observatory granted by the University of Hawaii, the University of California, and the California Institute of Technology. We thank the observers who contributed to the measurements reported here and acknowledge the efforts of the Keck Observatory staff. We extend special thanks to those of Hawaiian ancestry on whose sacred mountain of Mauna Kea we are privileged to be guests. SNIFS on the UH 2.2-m telescope is part of the Nearby Supernova Factory project, a scientific collaboration among the Centre de Recherche Astronomique de Lyon, Institut de Physique Nuclaire de Lyon, Laboratoire de Physique Nuclaire et des Hautes Energies, Lawrence Berkeley National Laboratory, Yale University, University of Bonn, Max Planck Institute for Astrophysics, Tsinghua Center for Astrophysics, and the Centre de Physique des Particules de Marseille. Based on data from the Infrared Telescope Facility, which is operated by the University of Hawaii under Cooperative Agreement no. NNX-08AE38A with the National Aeronautics and Space Administration, Science Mission Directorate, Planetary Astronomy Program. Some/all of the data presented in this paper were obtained from the Mikulski Archive for Space Telescopes (MAST). STScI is operated by the Association of Universities for Research in Astronomy, Inc., under NASA contract NAS5-26555. Support for MAST for non-HST data is provided by the NASA Office of Space Science via grant NNX09AF08G and by other grants and contracts. This research has made use of the NASA/IPAC Infrared Science Archive, which is operated by the Jet Propulsion Laboratory, California Institute of Technology, under contract with the National Aeronautics and Space Administration. This research made use of the SIMBAD and VIZIER Astronomical Databases, operated at CDS, Strasbourg, France (http://cdsweb.u-strasbg.fr/), and of NASAs Astrophysics Data System, of the Jean-Marie Mariotti Center Search service (http://www.jmmc.fr/searchcal), co-developed by FIZEAU and LAOG/IPAG. E.D.L. received funding from the European Union Seventh Framework Programme (FP7/2007- 2013) under grant agreement number 313014 (ETAEARTH). B.J.F. notes that this material is based upon work supported by the National Science Foundation Graduate Research Fellowship under grant No. 2014184874. Any opinion, findings, and conclusions or recommendations expressed in this material are those of the authors and do not necessarily reflect the views of the National Science Foundation.}

\bibliography{211351816v2}

\end{document}